\documentclass[useAMS,usenatbib]{mn2e}
\usepackage{graphicx,epsfig}
\usepackage{natbib}
\def\be{\begin{equation}}
\def\ee{\end{equation}}

\title{Stable chaos in the 55Cnc exoplanetary system?}

\author[J. Gayon, F. Marzari, H. Scholl]
{J.Gayon$^{1}$\thanks{E-mail:Julie.Gayon@oca.eu}, F. Marzari$^{2}$, H.Scholl$^{1}$\\
$^{1}$Laboratoire Cassiop\'ee, Universit\'e de Nice Sophia Antipolis, CNRS, Observatoire de la C\^ote d'Azur, B.P. 4229,
F-06304 Nice, France\\
$^{2}$Department of Physics, University of Padova, Via Marzolo 8,
35131 Padova, Italy\\
}

\begin{document}

\date{Draft version \today}

\maketitle

\begin{abstract}

The five planets discovered around the main-sequence star 55 Cnc 
may represent a case of stable chaos. By using both 
the Frequency Map Analysis and MEGNO we find that about 15 \% of the 
systems that can be build from the nominal orbital elements 
of the system are highly chaotic.  However, in spite of the 
fast diffusion rate in the phase space, the planetary system 
is not destabilized over 400 Myr and close encounters between the 
planets are avoided.  

\end{abstract}


\section{Introduction} \label{sec:intro}

The most crowded extrasolar planetary system discovered so far is 
that around the main-sequence star 55 Cnc(= $\rho^1$ Cancri).
Doppler shift measurements strongly suggest the existence of 
five planets orbiting the star with semimajor axes ranging from 
0.038 to 5.901 AU \citep{fima} . The innermost and smallest planet,
$e$, is a Neptune-mass object
while the outer and most massive one, planet $d$,  has a minimum mass 
of about 4 Jupiter masses.  A self-consistent dynamical fit to the 
stellar wobble data performed by \cite{fima} gives a set of 
orbital elements for the planets reported in Table 1. 
It is claimed \citep{fima} that this nominal system of five planets 
described in Table 1 is dynamically
stable at least over a timescale of 1 Myr. 
We would like to point out that planets $b$ and $c$ are no longer in a 
3:1 mean motion resonance as in the previous solution given by \cite{McA}. In
addition, the eccentricity of planet $c$ in \cite{fima} is significantly
smaller compared to that derived by \cite{McA}.

We have performed a detailed exploration of the stability of 
the nominal system by applying the Frequency Map Analysis (hereinafter FMA) 
method \citep{las1,las2,las3,masch} on 400 varied systems. These systems
are obtained by varying orbital elements of the nominal system. 
The varied systems
are analysed by adding randomly some inclination (lower than 5 deg)
to the Keplerian orbits to make the system more realistic.
The masses of the planets are scaled accordingly.
Among all the systems analysed with FMA we obtain 
cases with large diffusion speed in the phase space 
suggesting a chaotic evolution and possible instability. 
However, integrating a few of 
these chaotic systems over timescales of $10^8$ yrs, 
we never obtain destabilization. We, therefore, think that  
these chaotic system are
examples for stable chaos. This term was introduced by 
\cite{mino} to indicate the peculiar behaviour of 
asteroids with short Lyapunov times (of the
order of some $10^3$ yrs) which however show a 
remarkable stability over the age of the solar system \citep{mino}.
For the 55Cnc planetary system we find that in about 15\% of all cases 
the orbital elements of the four inner planets exhibit a
chaotic evolution on a timescale of a few million of years. 
This stable chaos may be due to the topology 
of the phase space: Quasi--periodic solutions in the
proximity of a chaotic island can act like ``quasi--barriers'' 
for the diffusion \citep{tsiva}.
The 
chaotic evolution appears to be confined 
and the system does not destabilize at least on a timescale of 
400 Myr. This peculiar behaviour may be due to the closeness of the
system to the planetary 3:1 mean motion resonance between planets 
$b$ and $c$ and secular 
resonances. 

To confirm that stable chaos is not limited to the
nominal case but can be retrieved also in a larger region of
the phase space, we randomly varied the orbital elements of 
the nominal case and found a similar behaviour. 
For a few selected cases, we also applied   
the MEGNO method (the acronym of Mean Exponential Growth
factor of Nearby Orbits) proposed by Cincotta \& Sim\`o (2000).
The results obtained by MEGNO confirm the FMA results.

\begin{table*}
\begin{center}   
   \caption{\label{table}Orbital parameters for the self-consistent dynamical
     fit of the 55 Cancri planetary systems. These data are taken from Table 4
     of Fischer et al. (2008).}
   \begin{tabular}{ccccccc}
   \hline
   Planet&
   $\begin{array}{c}
      \textrm{Period}\\
      \textrm{(days)}
   \end{array}$&
   $\begin{array}{c}
      T_p\\
      \textrm{(JD-2440000)}
   \end{array}$&
   $\begin{array}{c}
      e\\
      \textrm{} 
   \end{array}$&
   $\begin{array}{c}
      \omega \\
      \textrm{(deg)}
   \end{array}$&
   $\begin{array}{c}
      M\,\sin\, i\\
      (M_{Jup})
   \end{array}$&
   $\begin{array}{c}
      a \\
      \textrm{(AU)}
   \end{array}$
   \tabularnewline
   \hline

   b&
   $14.651262$&
   $7572.0307$&
   $0.0159$&
   $164.001$&
   $0.8358$&
   $0.115$
   \tabularnewline

   c&
   $44.378710$&
   $7547.5250$&
   $0.0530$&
   $57.405$&
   $0.1691$&
   $0.241$
   \tabularnewline

   d&
   $5371.8207$&
   $6862.3081$&
   $0.0633$&
   $162.658$&
   $3.9231$&
   $5.901$
   \tabularnewline

   e&
   $2.796744$&
   $7578.2159$&
   $0.2637$&
   $156.500$&
   $0.0241$&
   $0.038$
   \tabularnewline

   f&
   $260.6694$&
   $7488.0149$&
   $0.0002$&
   $205.566$&
   $0.1444$&
   $0.785$
   \tabularnewline
   \hline

\end{tabular}
\end{center}
\end{table*}

\begin{figure}
\includegraphics[width=8. cm, height=5 cm]{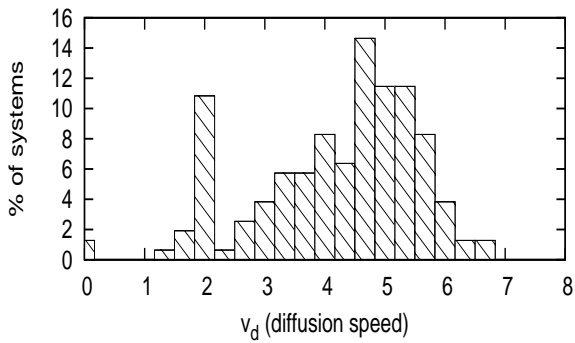}
\caption[]{Histogram showing the dsitribution of the diffusion rate 
within our sample
of 400 55Cnc planetary systems. 
}
\label{histo}
\end{figure}

\newpage
\section{FMA analysis}  \label{sec:FMA}

We have performed the FMA analysis on 400 different systems 
derived from Table 1. Random mean anomalies, node
longitudes and inclinations lower than $5^{\circ}$ are assigned to 
each planet. All other orbital elements were taken from Table 1. 
The frequency analysis is performed over a timescale of 
$2 \times 10^4$ yrs. The orbital elements computed with the numerical 
integrator SYMBA \citep{dun} are
Fourier analyzed and the 
values of intrinsic frequencies are obtained
over running windows. The relative changes of these frequencies
are estimated and used to compute the diffusion rate in the 
phase space. We have applied FMA to the signal $s_{\Delta\varpi_{b,c}}$,
the difference between the periapse longitude of planets $b$ and $c$, 
and to the more conventional signal $s = h_e + i k_e$. The usual 
non--singular variables $h_e = ecc*cos(\varpi)$ and $k_e = ecc*sin(\varpi)$
refer to the innermost planet $e$ where ecc refers to its eccentricity.
The diffusion rate is computed by using the standard deviation $\sigma$ of the 
main frequency of the signal computed over the running windows. 
Slow diffusion rates, characterized by small values of $\sigma$, 
mean quasi--periodic systems, while large values of $\sigma$ 
imply chaotic evolution. As in \cite{masch} we measure the diffusion
speed by the logarithmic number $v_d = -log_{10}(\sigma) + log_{10}(\sigma_0)$
where $\sigma_0$ is the smallest value of $\sigma$ we observed in our
sample of systems.
A small value of $v_d$ means a low dispersion of the frequencies in the 
considered time interval and, hence,  a slow diffusion speed in the phase
space. 
Large values of $v_d$ indicate fast changes of the system frequencies
and, therefore, chaos. 

In Fig.\ref{histo} we show the distribution of 
the diffusion speed measured by $v_d$ 
in our sample of 400 55Cnc planetary systems. 
Values smaller than 4 suggestlong term stability according to
our previous experience with FMA, while
larger values indicate chaos. We concentrate on systems with 
a diffusion speed of about 5 or larger which  represent about 15\% of the 
whole sample we analysed. These systems 
have 
major frequencies which change
on short timescales. If the fast diffusion of major frequencies induces
also drastic changes of amplitudes, in particular of planetary eccentricities,
close encounters between planets may occur resulting in the ejection of 
one or more planets.  We will show in the next section  that planetary eccentricities
do not increase on longer timescales  which suggets a ``stable chaos'' state
for the Cnc55 system.

An additional set of FMA simulations has been performed for 400 systems 
where we have randomly varied the last signifcant digit of all the orbital elements 
given in Table 1. We obtain a histogram that is very similar
to that shown in  Fig.\ref{histo} confirming that the behaviour of
stable chaos we have found
for the nominal case of Table 1 is extended in phase space. 

\section{Long term evolution of the chaotic systems} \label{sec:model}

\begin{figure}
\hskip -1.2 truecm
\includegraphics[width=14.0 cm]{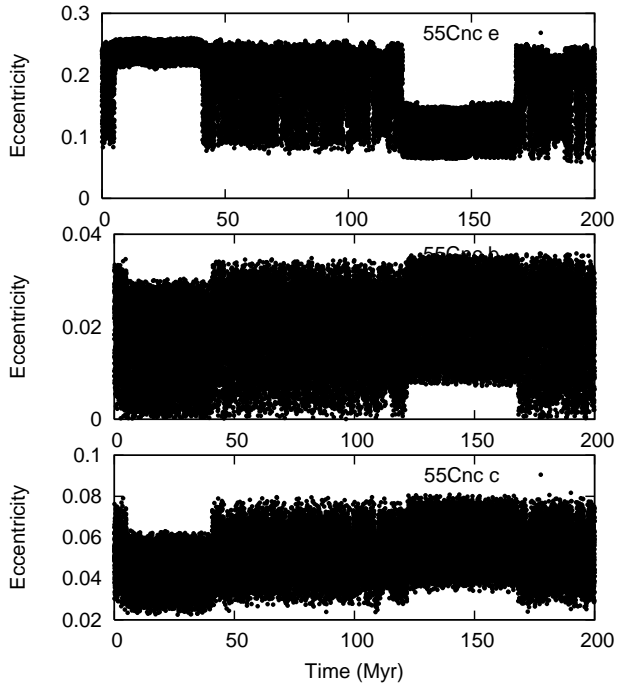}
\caption[]{Evolution with time of the eccentricity of 
the three inner planets of the system. The top plot 
shows the eccentricity of 55Cnc e, the medium plot that 
of 55Cnc b and the bottom plot that of 55Cnc c.
}
\label{long_ecc}
\end{figure}

For those systems with a large diffusion speed ($v_d >4$) 
we performed numerical integrations over a timescale of 400 Myr
with SYMBA \citep{dun}. 
In Fig.\ref{long_ecc} we show for a typical run
the evolution of the eccentricity of the 3 inner planets $e,b,c$. The eccentricity
jumps reveal the chaotic nature of the system. The jumps are obviously correlated
among the three planets. Major jumps occur simultaneously. Also the eccentricity of
planet $f$ shows simultaneously
jumps but more moderate. All jumps, although they appear
to be significant, do not result in close approaches among the planets. 
The planetary system is not destabilized on this timescale while it is chaotic.

Fig.\ref{long_ecc} suggests that 
we have the case of stable chaos. The eccentricity evolution is  
characterized by jumps.
It does not have a random walk growth but remains limited. 
It appears that the system is bouncing between limited regions.

\begin{figure}
\hskip -1.2 truecm
\includegraphics[width=14.0 cm]{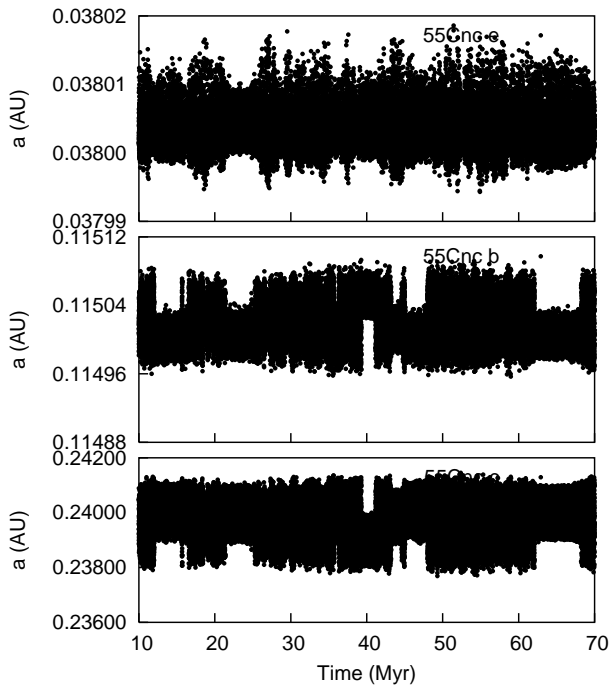}
\caption[]{Semimajor axis of the three inner planets of the 
system over a short timescale. 
}
\label{sma}
\end{figure}

A different case with similar high 
diffusion speed $v_d$  is shown in Fig.\ref{sma}. The semimajor axes of the 
three inner planets are shown on a shorter timescale to highlight the 
chaotic evolution. Also in this case the system remains confined in 
a stable configuration over 400 Myr. 

\section{MEGNO ANALYSIS}

Besides the FMA method which uses the diffusion rate of intrinsic frequencies
to measure chaos, there is a large class of methods which use the divergence
of nearby orbits in phase space as a measure. Exponential divergence means
chaos while linear divergence means non-chaotic which guarantees stability 
of the dynamical system. Stability means here that, in particular, no planet is ejected out
of the system. A nearly exponential divergence does not necessarily mean
instability in our sense over the lifetime of the system. As outlined above,
we would qualify such a system as stable chaotic. 
The widely used Lyapunov characteristic numbers (LCNs) yield the necessary information
about the divergence and are used to measure the chaocity of the system. 
Since the computation of
the LCNs is very time consuming, more rapid methods were introduced which
were shown to yield a very good estimation for the LCNs. One of these fast
LCN estimators is the MEGNO \citep{cisi} method. It was applied to several exoplanetary
systems \citep{bois20023}.
We computed the MEGNO indicator for about 50 varied systems. They were obtained
by varying only the eccentricites of planets e,b,c and f 
while taking the other orbital elements from the nominal system
of Table 1. Eccentricites of planet
$e$ are varied between 0.22 and 0.26, of planets $b$ and $c$ between 0. and 0.10
and of planet $f$ between 0. and 0.04. We integrated the systems 
over 100 000 yrs. For eccentricities of planets $b$ and $c$
ranging from 0. to 0.02, we obtain weak chaos. The estimator for the maximal
Lyapunov exponent is not linear but far from exponential. For larger eccentricities,
the indicator has a behaviour between linear and exponential.
MEGNO shows, like
FMA, that the four planetary orbits close to the nominal system are chaotic.

\section{Conclusions} \label{sec:conclusions}

Among the possible dynamical configurations of the 55Cnc planetary 
system there is also stable chaos. About 15\% of the systems built from the 
nominal dynamical fit given in \cite{fima} have a high diffusion 
speed in the phase space. However, the chaotic island seems to 
be limited in extent and surrounded by quasi--periodic systems. 
In this way, the chaotic systems do not increase their eccentricity
to planet crossing values and they are stable over a long timescale. 
Systems with intermediate values of diffusion speed $v_d$ 
as measured by FMA ($\sim 4$) do not show 
this behaviour. They possibly populate the outer border of the chaotic
region and their chaotic evolution is much slower. Systems with 
smaller values of  $v_d$ are possibly stable over timescale of $10^9$ yrs.
Additional FMA and MEGNO computations show that this behaviour is 
not only peculiar for the nominal system given by Table 1 
but it is extended over a wider range of orbital elements for the system. 
The weak resonances responsible for the stable chaos are present also
for different combinations of the initial orbital elements.  

\section{Acknowledgments} 
We thank J. Hadjidemetriou for his useful comments and suggestions that helped 
to improve the manuscript.
The MEGNO computations were performed on the "Mesocentre SIGAMM" machine,
hosted by the Observatoire de la C\^ote d'Azur.

{}
\clearpage


\begin{thebibliography}{}
%
\bibitem[\protect\citeauthoryear{Bois et al.}{2003}]{bois20023} Bois, E.,
Kiseleva-Eggleton, L., Rambaux, N., 
Pilat-Lohinger, E., 2003, ApJ 598, 1312-1320
%
\bibitem[\protect\citeauthoryear{Cincotta \& Sim\`o}{2000}]{cisi} 
Cincotta, P., \& Sim\'o, C., 2000, A\&AS 147, 205-228
%
\bibitem[\protect\citeauthoryear{Duncan et al.}{1998}]{dun} 
Duncan, M. J., Levison, H. F., Lee, M.H., 1998,  AJ 116, 2067-2077
%
\bibitem[\protect\citeauthoryear{Fischer et al.}{2008}]{fima} 
Fischer, D. A., Marcy, G. W., Butler, R. P., Vogt, S. S., Laughlin, G., 
Henry, G. W., Abouav, D., Peek, K. M. G., Wright, J. T., Johnson, J. A., 
McCarthy, C., Isaacson, H., 2008, ApJ 675, 790-801
%
\bibitem[\protect\citeauthoryear{Laskar et al.}{1992}]{las1} 
Laskar J., Froeschl\'e C., Celletti A., 1992, Physica D 56, 253-269
%
\bibitem[\protect\citeauthoryear{Laskar}{1993a}]{las2}  
Laskar J. 1993, Physica D 67, 257-281
%
\bibitem[\protect\citeauthoryear{Laskar}{1993b}]{las3} 
Laskar J. 1993, Cel. Mech. and Dyn. Astr.  56,  191-196
%
\bibitem[\protect\citeauthoryear{Marzari et al.}{2006}]{masch} 
Marzari, F., Scholl, H., Tricarico, P., 2006, A\&A 453, 341-348
%
\bibitem[\protect\citeauthoryear{McArthur et al.}{2004}]{McA} 
McArthur, B. E., Endl, M., Cochran, W. D., Benedict, G. F., Fischer, D. A., 
Marcy, G. W., Butler, R. P., Naef, D., Mayor, M., Queloz, D., Udry, S.,
Harrison, T. E., 2004, ApJ 614, L81-L84
%
\bibitem[\protect\citeauthoryear{Milani \& Nobili}{1992}]{mino} 
Milani, A., Nobili, A., 1992, Nature 357, 569-571
%
\bibitem[\protect\citeauthoryear{Tsiganis et al.}{2000}]{tsiva} 
Tsiganis, K., Varvoglis, H., Hadjidemetriou, D., 2000, Icarus 146, 240-252
%
\end{thebibliography}
\end{document}